# Applying and Combining Three Different Aspect Mining Techniques

M. Ceccato, M. Marin, K. Mens, L. Moonen,
P. Tonella, and T. Tourwé

TUDelft

SERG

# Applying and Combining Three Different Aspect Mining Techniques

M. Ceccato[1], M. Marin[2], K. Mens[3], L. Moonen[2,4], P. Tonella[1], and T. Tourwé[4]

[1] ITC-irst, Trento, Italy
[2] Delft University, The Netherlands
[3] Université catholique de Louvain, Belgium
[4] CWI, The Netherlands

`ceccato@itc.it, a.m.marin@ewi.tudelft.nl, kim.mens@uclouvain.be,`
`leon.moonen@computer.org, tonella@itc.it, tom.tourwe@cwi.nl`

**Abstract.** Understanding a software system at source-code level requires understanding the different concerns that it addresses, which in turn requires a way to identify these concerns in the source code. Whereas some concerns are explicitly represented by program entities (like classes, methods and variables) and thus are easy to identify, *crosscutting* concerns are not captured by a single program entity but are *scattered* over many program entities and are *tangled* with the other concerns. Because of their crosscutting nature, such crosscutting concerns are difficult to identify, and reduce the understandability of the system as a whole.

In this paper, we report on a combined experiment in which we try to identify crosscutting concerns in the JHotDraw framework automatically. We first apply three independently developed aspect mining techniques to JHotDraw and evaluate and compare their results. Based on this analysis, we present three interesting combinations of these three techniques, and show how these combinations provide a more complete coverage of the detected concerns as compared to the original techniques individually. Our results are a first step towards improving the understandability of a system that contains crosscutting concerns, and can be used as a basis for refactoring the identified crosscutting concerns into aspects.

## 1 Introduction

The increasing popularity of aspect-oriented software development (AOSD) is largely due to the fact that it recognises that some concerns cannot be captured adequately using the abstraction mechanisms provided by traditional programming languages. Several examples of such *crosscutting* concerns have been identified, ranging from simple ones such as logging, to more complex ones such as transaction management [1] and exception handling [2, 3].

An important problem with such crosscutting concerns is that they affect the understandability of the software system, and as a result reduce its evolvability and maintainability. First of all, crosscutting concerns are difficult to understand, because their implementation can be scattered over many different packages, classes and methods. Second, in the presence of crosscutting concerns, ordinary concerns become harder to understand as well, because they get tangled with the crosscutting ones: particular classes and methods do not only deal with the primary concern they address, but also may need to take into account some secondary, crosscutting concerns.

Several authors have presented automated code mining techniques, generally referred to as *aspect mining* techniques, that are able to identify crosscutting concerns in the source code [4]. The goal of these techniques is to provide an overview of the source-code entities that play a role in a particular crosscutting concern. This not only improves the understandability of the concern in particular and of the software in general, but also provides a first step in the migration towards applying aspect-oriented software development techniques. However, since the research field is still in its infancy, very few experiments have been conducted on real-world case studies, comparisons of different techniques are lacking, and no agreed-upon benchmark is available that allows to evaluate the existing techniques.

This paper reports on an experiment involving three independently developed aspect mining techniques: fan-in analysis [5, 6], identifier analysis [7, 8] and dynamic analysis [9]. In the experiment, each of these techniques is applied to the same case study: the JHotDraw graphical editor framework. The goal of the experiment is not to identify the "best" aspect mining technique, but rather to mutually compare the individual techniques and assess their major strengths and weaknesses. Additionally, by identifying where the techniques overlap and where they are complementary, the experiment allows us to propose interesting combinations and to apply these combinations on the same benchmark to verify whether they actually perform better.

The JHotDraw framework which we selected as benchmark case was originally developed to illustrate good use of object-oriented design patterns [10] in Java programs. This implies that the case study has been well-designed and that care has been taken to cleanly separate concerns and make it as understandable as possible. Nevertheless, JHotDraw exposes some of the modularisation limitations present even in well-designed systems, and contains some quite interesting crosscutting concerns.

The contributions of this paper can be summarised as follows:

- We provide an overview of the major strengths and weaknesses of three aspect mining techniques. This information is valuable for developers using these techniques, as it can help them choosing a technique that suits their needs. Other aspect mining researchers can take this information into account to compare their techniques to ours, or to fine-tune our techniques;
- We discuss how the individual techniques can be combined in order to perform better, and validate whether this is indeed the case by applying the combined techniques on the same benchmark application and comparing the results;
- We present a list of all crosscutting concerns that the three techniques identified in the JHotDraw framework. Such information is valuable for other

```
interface A {
  public void m();
}
class B implements A {
  public void m() {};
}
class C1 extends B {
  public void m() {};
}
class C2 extends B {
  public void m() { super.m();};
}
class D {
  void f1(A a) { a.m(); }
  void f2(B b) { b.m(); }
  void f3(C1 c) { c.m(); }
}
```

**Fig. 1.** Various (polymorphic) method calls.

aspect mining researchers who want to validate their techniques, and might lead to JHotDraw becoming a de-facto benchmark for aspect mining techniques;

The paper is structured as follows. Section 2 introduces the necessary background concepts required to understand the three aspect mining techniques explained in Section 3. Section 4 presents the results of applying each technique on the common benchmark, while Section 5 uses these results for discussing the benefits and drawbacks of each technique with respect to the others. Based on this discussion, Section 6 presents useful combinations of the techniques, and reports on the experience of applying such combinations on the benchmark application. Section 7 presents our conclusions. For an overview of related work concerning aspect mining, we refer to the papers discussing the individual techniques [5–9] and to an initial survey on aspect mining [4].

## 2 Background concepts

### 2.1 Fan-in

The *fan-in* metric, as defined by Henderson-Sellers, counts the number of locations from which control is passed into a module [11]. In the context of object-orientation, the module-type to which this metric is applied is the method. We define the *fan-in* of a method $M$ as the number of distinct method bodies that can invoke $M$. Because of polymorphism, one call site can affect the fan-in of several methods: a call to method $M$ contributes to the fan-in of $M$, *but also* to all methods refined by $M$, *as well as* to all methods that are refining $M$ [6].

| Method | Potential callers | Fan-in |
|--------|-------------------|--------|
| A.m | D.f1, D.f2, D.f3 | 3 |
| B.m | D.f1, D.f2, D.f3, C2.m | 4 |
| C1.m | D.f1, D.f2, D.f3 | 3 |
| C2.m | D.f1, D.f2 | 2 |

**Fig. 2.** Fan-in values for program in Figure 1.

As an example, Figure 2 shows the calculated fan-in for the methods named $m$ in the program of Figure 1. Note that $D.f3$ is reported among the potential callers of $B.m$, even though this situation cannot actually occur at run-time. However, the resulting effect of having higher fan-in values reported for methods in super-classes is arguably positive for the purpose of the present analysis, as it emphasizes the concern implemented by the super-class method, which generally is addressed by its overriding methods as well.

## 2.2 Concept analysis

Formal concept analysis (FCA) [12] is a branch of lattice theory that can be used to identify meaningful groupings of *elements* that have common *properties*.[5]

| Programming lang. | object-oriented | functional | logic | static typing | dynamic typing |
|-------------------|-----------------|------------|-------|---------------|----------------|
| Java | √ | - | - | √ | - |
| Smalltalk | √ | - | - | - | √ |
| C++ | √ | - | - | √ | - |
| Scheme | - | √ | - | - | √ |
| Prolog | - | - | √ | - | √ |

**Table 1.** Programming languages and their supported programming paradigms.

FCA takes as input a so-called *context*, which consists of a (potentially large, but finite) set of *elements* $E$, a set of *properties* $P$ on those elements, and a Boolean *incidence relation* $T$ between $E$ and $P$. An example of such a context is given in Table 1, which relates different programming languages and properties. A mark √ in a table cell means that the element (programming language) in the corresponding row has the property of the corresponding column.

Starting from such a context, FCA determines *maximal* groups of elements and properties, called *concepts*, such that each element of the group shares the properties, every property of the group holds for all of its elements, no other

---

[5] We use the terms *element* and *property* instead of *object* and *attribute* used in traditional FCA literature, because these latter terms have a very specific meaning in OO software development.

                                                                                                    

element outside the group has those same properties, nor does any property outside the group hold for all elements in the group. Intuitively, a *concept* corresponds to a maximal 'rectangle' containing only $\sqrt{}$ marks in the table, modulo any permutation of the table's rows and columns.

Formally, the starting context is a triple $(E, P, T)$, where $T \subseteq E \times P$ is a binary relation between the set of all elements $E$ and the set of all considered element properties $P$. A *concept $c$* is defined as a pair of sets $(X, Y)$ such that:

$$X = \{e \in E \mid \forall p \in Y : (e, p) \in T\} \tag{1}$$

$$Y = \{p \in P \mid \forall e \in X : (e, p) \in T\} \tag{2}$$

where $X$ is said to be the *extent* of the concept ($Ext[c]$) and $Y$ is said to be its *intent* ($Int[c]$). It should be noticed that the definition above is not "constructive", being mutually recursive between $X$ and $Y$. However, given a pair $(X, Y)$, it allows deciding whether it is a concept or not. FCA algorithms provide constructive methods to determine all pairs $(X, Y)$ satisfying the constraints (1) and (2).

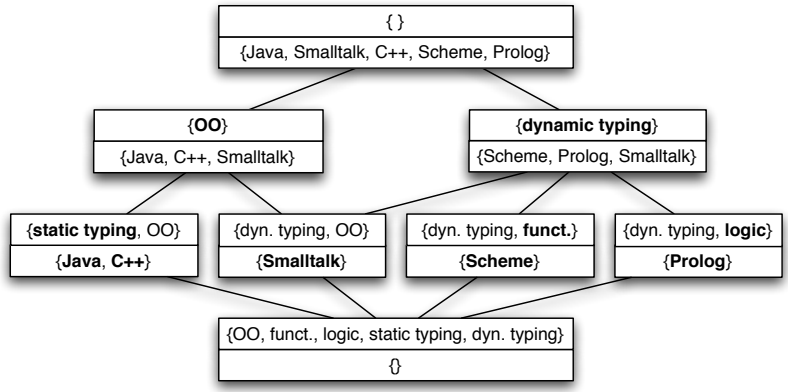

**Fig. 3.** The concept lattice for Table 1.

The containment relationship between concept extents (or, equivalently, intents) defines a partial order over the set of all concepts, which can be shown to be a lattice [12]. Figure 3 shows the concept lattice corresponding to Table 1. The lattice's bottom concept contains those elements that have all properties. Since there is no such programming language in our example, that concept contains no elements (its extent is empty). Similarly, the top concept contains those properties that hold for all elements. Again, there is no such property (the concept's intent is empty). Other concepts represent related groups of programming languages, such as the concept ({*Java, C++*}, {*static typing, OO* }), which groups all statically-typed object-oriented languages, a sub-concept of all OO languages. Intuitively, the sub-concept relationship can thus be interpreted as a specialization of more general notions. Elements (resp. properties) in boldface are those that are most concept-specific, being attached to the largest lower bound (resp. least upper bound) concept. When using the so-called *sparse labeling* of the concept lattice, only these boldface labels are retained, without loss of information.

More precisely, when using *sparse labeling*, a node $c$ is marked with an element $e \in Ext[c]$ only if it is associated with the most specific (i.e., lowest) concept $c$ having $e$ in the extent; a node $c$ is marked with a property $p \in Int[c]$ only if it is associated with the most general (i.e., highest) concept $c$ having $p$ in its intent. The (unique) node of a lattice $L$ marked with a given element $e$ is thus:

$$\gamma(e) = \inf\{c \in L \mid e \in Ext[c]\} \qquad (3)$$

where *inf* gives the infimum (largest lower bound) of a set of concepts. Similarly, the unique lattice node marked with a given property $p$ is:

$$\mu(p) = \sup\{c \in L \mid p \in Int[c]\} \qquad (4)$$

where *sup* gives the supremum (least upper bound) of a set of concepts. The set of elements in the extent of a lattice node $c$ can then be computed as the set of all elements at or below $c$, while the set of properties in its intent are those marking $c$ or any node above $c$.

The labeling introduced by the functions $\mu$ and $\gamma$ give the most specific concept for a given element (resp. property). Thus, with sparse labeling, the elements and properties that label a given concept are those that characterize it most specifically. Sometimes it is convenient to get the labels of a given concept through the following functions:

$$\alpha(c) = \{p \in P \mid \mu(p) = c\} \qquad (5)$$

$$\beta(c) = \{e \in E \mid \gamma(e) = c\} \qquad (6)$$

$\alpha(c)$ gives the set of properties labeling a concept $c$, while $\beta(c)$ gives the concept's elements, according to the *sparse labeling*.

## 2.3 Terminology

We conclude this background section by introducing some terminology that will be used throughout the remainder of this paper.

**A concern** is a collection of related source-code entities, such as classes, methods, statements or expressions, that implement a particular functionality or feature of the application. A *crosscutting* concern is a concern whose entities are not captured into a single localised abstraction, but are scattered over many different locations and tangled with other concerns.

**A (concern) seed** is a single source-code entity, such as a method, or a collection of such entities, that strongly connotes a crosscutting concern. It offers a starting point for further exploration and understanding the whole extent of that concern's implementation.

**A candidate seed** is identified by an automated aspect mining technique as a potential concern seed but is not yet confirmed to be an actual concern seed or rather a false positive.

**Seed expansion** is the manual or automated process of completing the set of source-code entities constituting a seed into the entire set of source-code entities of which the crosscutting concern corresponding to that seed consists.

## 3   The three aspect mining techniques

In this section, we give a brief overview of three techniques, developed independently by different research groups, that support the automated discovery of crosscutting concerns in the source code of a software system that is written in a non aspect-oriented way.

### 3.1   Fan-in analysis

Crosscutting functionality can occur at different levels of modularity. Classes, for instance, can assimilate new concerns by implementing multiple interfaces or by implementing new methods specific to super-imposed roles. At the method level, crosscutting in many cases resides in calls to methods that address a different concern than the core logic of the caller. Typical examples include logging, tracing, pre- and post-condition checks, and exception handling. It is exactly this type of crosscutting that fan-in analysis tries to capture.

When we study the mechanics of AOSD, we see that it employs the so-called *advice* construct to eliminate crosscutting at method level. This construct is used to acquire control of program execution and to add crosscutting functionality to methods without an explicit invocation from those methods. Rather, the crosscutting functionality is isolated in a separate module, called aspect, and woven with the method implicitly based on the advice specification.

Fan-in analysis reverses this line of reasoning and looks for crosscutting functionality that is explicitly invoked from many different methods scattered throughout the code. The hypothesis is that the *amount* of calls to a method implementing this crosscutting functionality (fan-in) is a good measure for the importance and scattering of the discovered concern.

To perform the fan-in analysis, a fan-in metric was implemented as a plug-in for the Eclipse platform[6], and integrated it into an iterative process that consists of three steps:

1. Automatic computation of the fan-in metric for all methods in the investigated system.
2. Filtering of the results from the previous step by
   - eliminating all methods with fan-in values below a chosen threshold (in the experiment, a threshold of 10 was used);

---

[6]  http://swerl.tudelft.nl/view/AMR/FINT

    – eliminating the accessor methods (methods whose signature matches a *get\*/set\** pattern and whose implementation only returns or sets a reference);

    – eliminating utility methods, like `toString()` and collection manipulation methods, from the remaining subset.

3. (Partially automated) analysis of the methods in the resulting, filtered set by exploring the callers, call sites, naming convention used, the implementation and the comments in the source code.

   Besides code exploration, the tool supports automatic recognition of a number of relations between the callers of a method, such as common roles, consistent call positions, etc.

The result of the fan-in analysis is a set of candidate seeds, represented as methods with high fan-in.

## 3.2 Identifier analysis

In the absence of designated language constructs for aspects, naming conventions are the primary means for programmers to associate related but distant program entities. This is especially the case for object-oriented programming, where polymorphism allows methods belonging to different classes to have the same signature, where it is good practice to use intention-revealing names [13], and where design and other programming patterns provide a common vocabulary known by many programmers.

*Identifier analysis* relies on this assumption and identifies candidate seeds by grouping program entities with similar names. More specifically, it applies FCA with as elements all classes and methods in the analyzed program (except those that generate too much noise in the results, like test classes and accessor methods), and as properties the identifiers associated with those classes and methods.

The identifiers associated with a method or class are computed by splitting up its name based on where capitals appear in it. For example, a method named `createUndoActivity` yields three identifiers `create`, `undo` and `activity`. In addition, we apply the Porter stemming algorithm [14] to make sure that identifiers with the same root form (like `undo` and `undoable`) are mapped to one single representative identifier or 'stem'. It is these stems that are used as properties for the concept analysis.

The FCA algorithm then groups entities with the same identifiers. When such a group contains a certain minimum number of elements (in the experiment, a threshold of 4 was used) and the entities contained in it cut across multiple class hierarchies, the group is considered a candidate seed. The only remaining but most difficult task is that of deciding manually whether a candidate seed is a real seed or a false positive. To help the developer in this last task, the *DelfSTof* source-code mining tool presents the concepts in such a way that they can be browsed easily by a software engineer and so that he or she can readily access the code of the classes and methods belonging to a discovered seed.

### 3.3   Dynamic analysis

Formal concept analysis has been used to locate 'features' in procedural programs [15]. In that work, the goal was to identify the computational units (procedures) that specifically implement a feature (i.e., requirement) of interest. Execution traces obtained by running the program under given scenarios provided the input data (dynamic analysis).

In a similar way, dynamic analysis can be used to locate aspects in program code [9] according to the following procedure. Execution traces are obtained by running an instrumented version of the program under analysis, for a set of scenarios (use-cases). The relationship between execution traces and executed computational units (methods) is subjected to concept analysis. The execution traces associated with the use-cases are the elements of the concept analysis context, while the executed methods are the properties. In the resulting concept lattice (with sparse labeling), the *use-case specific* concepts are those labeled by at least one trace for some use-case (i.e. $\alpha$ contains at least one element), while the concepts with zero or more properties as labels (those with an empty $\alpha$) are regarded as *generic* concepts. Thus, use-case specific concepts are a subset of the generic ones.

Both use-case specific concepts and generic concepts carry information potentially useful for aspect mining, since they group specific methods that are always executed under the same scenarios. When the methods that label one such concept (using the sparse labeling) crosscut the principal decomposition, a candidate aspect is determined.

Formally, let $C$ be the set of all the concepts and let $C_s$ be the set of use-case specific concepts ($|\alpha(c)| > 0$). A concept $c$ is considered a candidate seed *iff*:

**Scattering:** $\exists p, p' \in \beta(c) \mid pref(p) \neq pref(p')$
**Tangling:** $\exists p \in \beta(c), \exists c' \in \Omega, \exists p' \in \beta(c') \mid c \neq c' \wedge pref(p) = pref(p')$

where $\Omega = C_s$ for the *use-case specific* seeds, while $\Omega = C$ for the *generic* seeds. The first condition (*scattering*) requires that more than one class contributes to the functionality associated with the given concept ($\texttt{pref}(p)$ is the fully scoped name of the class containing the method $p$). The second condition (*tangling*) requires that the same class addresses more than one concern.

In summary, a concept is a candidate seed if: (1) *scattering:* more than one class contributes to the functionality associated with the given concept; (2) *tangling:* the class itself addresses more than one concern.

The first condition alone is typically not sufficient to identify crosscutting concerns, since it is possible that a given functionality is allocated to several modularized units without being tangled with other functionalities. In fact, it might be decomposed into sub-functionalities, each assigned to a distinct module. It is only when the modules specifically involved in a functionality contribute to other functionalities as well (i.e. the second condition) that crosscutting is detected, hinting for a candidate seed.

| Concern type | # | Seed's description |
|---|---|---|
| Consistent behavior | 4 | Methods implementing the consistent behavior shared by different callers, such as checking and refreshing figures/views that have been affected by the execution of a command. |
| Contract enforcement | 4 | Method implementing a contract that needs to be enforced, such as checking the reference to the editor's active view before executing a command. |
| Undo | 1 | Methods checking whether a command is undoable/redoable and the *undo* method in the superclass, which is invoked from the overriding methods in subclasses. |
| Persistence and resurrection | 1 | Methods implementing functionality common to persistent elements, such as read/write operations for primitive types wrappers (e.g., Double, Integer, etc.) which are referenced by the scattered implementations of persistence/resurrection. |
| Command design pattern | 1 | The *execute* method in the command classes and command constructors. |
| Observer design pattern | 1 | The observers' manipulation methods and *notify* methods in classes acting as subject. |
| Composite design pattern | 2 | The composite's methods for manipulating child components, such as adding a new child. |
| Decorator design pattern | 1 | Methods in the decorator that pass the calls on to the decorated components. |
| Adapter design pattern | 1 | Methods that manipulate the reference from the adapter (*Handle*) to the adaptee (*Figure*). |

**Table 2.** Summary of the results of the fan-in analysis experiment.

## 4 Results of the aspect mining

In this section, we present the results of applying each technique to version 5.4b1 of JHotDraw, a Java program with approximately 18,000 non-commented lines of code and around 2800 methods. We mutually compare the results of the techniques, and discuss the limitations of each technique as well as their complementarity.

### 4.1 The fan-in analysis experiment

As described in Subsection 3.1, fan-in analysis first performs a number of successive steps to filter the methods in the analyzed system. The threshold-based filtering, which selects methods with high fan-in values, kept around 7% of the total number of methods. The filters for accessors and utility methods eliminated around half of the remaining methods. In the remaining subset, more than half of the methods (52%) were categorized as seeds, based on manual analysis.

Table 2 gives an overview of the types of crosscutting concerns that were identified and the seeds that led to their identification. Several of these concern types, such as *consistent behavior* or *contract enforcement* [16], have more than one instance in JHotDraw; that is, multiple unrelated (crosscutting) concerns exist that conform to the same general description. For example, one instance of *contract enforcement* checks a priori conditions to a command's execution, while another instance verifies common requirements for activating drawing tools. The number of different instances that were detected is indicated in the # column.

We distinguish three different ways in which the fan-in metric can be associated with the crosscutting structure of a concern implementation (also indicated in Table 2):

1. The crosscutting functionality is implemented through a method and the crosscutting behavior resides in the explicit calls to this method. Examples in this category include *consistent behavior* and *contract enforcement*.
2. The implementation of the crosscutting concern is scattered throughout the system, but makes use of a common functionality. The crosscutting resides in the call sites, and can be detected by looking at the similarities between the calling contexts and/or the callers. Examples of concerns in this category are *persistence* and *undo* [6].
3. The methods reported by the fan-in analysis are part of the roles superimposed to classes that participate in the implementation of a design pattern. Many of these roles have specific methods associated to them: the *subject* role in an Observer design pattern is responsible to notify and manage the observer objects, while the *composite* role defines specific methods for manipulating child components. In general, establishing a relation between these seed-methods and the complete concern to which they appertain might require a better familiarity of the human analyzer with the code being explored, than for the previous two categories. However, many of these patterns are well-known and have a clear defined structure, which eases their recognition [17].

For more details regarding fan-in analysis and a complete discussion of the JHotDraw results, we refer to [6].

## 4.2 The identifier analysis experiment

Applying the identifier analysis technique of Subsection 3.2 on JHotDraw yielded 230 concepts and took about 31 seconds when using a threshold of 4 for the minimum number of elements in a concept. With a threshold of 10, the number of concepts produced was significantly less: only 100 concepts remained after filtering, for a similar execution time.[7] In both cases, 2193 elements and 507 properties were considered. It is a good sign that the number of properties is significantly smaller than the total number of elements considered, as it implies that there is quite some overlap in the identifiers of the different source-code entities, which was one of the premisses of the identifier analysis technique.

---

[7] Whereas the threshold of 4 was chosen arbitrarily, the threshold of 10 was determined experimentally: below that threshold the amount of concepts that were regarded as noise was significantly higher than above the threshold.

| Crosscutting concern | Concept(s) | #elements | Some elements |
|---|---|---|---|
| Observer | change(d) | 67 | figureChanged(e) |
| | check | 14 | checkDamage() |
| | listener | 65 | createDesktopListener() |
| | release | 12 | . . . |
| Command execution | command executed | 4 | commandExecuted(...) |
| | execut(abl)e | 51 | commandExecutable(...) |
| Undo | undo(able) | 53 | createUndoActivity() |
| | redo(able) | 14 | redo() |
| Visitor | visit | 12 | visit(FigureVisitor) |
| Persistence | file | 15 | registerFileFilters(c) |
| | storable | 5 | readStorable() |
| | load | 8 | loadRegisteredImages |
| | register | 7 | loadRegisteredImages |
| Drawing figures | draw | 112 | draw(g) |
| Moving figures | move | 36 | moveBy(x,y) |
| | | | moveSelection(dx,dy) |
| Iterating over collections | iterator | 5 | iterator(), listIterator(), . . . |

**Table 3.** Selection of results of the identifier analysis experiment.

The manual part of the experiment, i.e. deciding which concepts were real seeds, was much more time-consuming. Overall, this took about three days for the experiment with threshold 4, where 230 seed candidates needed to be investigated. For each of the discovered concepts, the code of the entities in its extent had to be inspected to decide whether (most of) these entities addressed a similar concern. Other than allowing to browse the source code of the elements in the extent of a concept, the DelfSTof code mining tool provided no direct support for this.

Table 3 presents some of the seeds discovered by manually analyzing the classes and methods belonging to the extent of the concepts produced by the FCA algorithm. The first column names the concern, the second column shows the identifiers shared by the elements belonging to the concept(s) corresponding to that concern. The third column shows the size of the extent for each concept. Finally, for illustration purposes, the fourth column shows some program entities appearing in the extent of the discovered concepts.

Out of 230 candidate seeds, 41 seeds were retained, when using a threshold of 4 for the minimum number of elements in a concept. These discovered concerns were classified in three different categories:

1. Some of these concerns looked like aspects in the more traditional sense (e.g., *observer*, *undo* and *persistence*).

2. Many other concerns seemed to represent a crosscutting functionality that was part of the business logic (e.g., *drawing figures*, *moving figures*). The distinction between these two first categories was somewhat subjective, however.

3. Three Java-specific concerns were discovered (e.g., *iterating over collections*) that are difficult to factor out into an aspect because they rely on or extend specific Java code libraries.

## 4.3   The dynamic analysis experiment

The dynamic analysis technique of Subsection 3.3 is supported by the *Dynamo* aspect mining tool[8]. The first step required by *Dynamo* is the definition of a set of use-cases. To accomplish this task, the documentation associated with the main functionalities of JHotDraw was used to define a use-case for each functionality described in the documentation. Amongst others, a use-case was created to draw a rectangle, one to draw a line using the scribble tool, one to create a connector between two existing figures, one to attach a URL to a graphical element, and so on. In total, 27 use-cases were obtained. When executed they exercised 1262 methods belonging to JHotDraw classes, so that the initial context for the concept analysis algorithm contained 27 elements and 1262 properties. The resulting concept lattice contained 1514 nodes.

| Crosscutting concern | Concepts | Methods |
|---|---|---|
| Undo | 2 | 36 |
| Bring to front | 1 | 3 |
| Send to back | 1 | 3 |
| Connect text | 1 | 18 |
| Persistence | 1 | 30 |
| Manage handles | 4 | 60 |
| Manage figure change event | 3 | 8 |
| Move figure | 1 | 7 |
| Command executability | 1 | 25 |
| Connect figures | 1 | 55 |
| Figure observer | 4 | 11 |
| Add text | 1 | 26 |
| Add URL to figure | 1 | 10 |
| Manage figures outside drawing | 1 | 2 |
| Get attribute | 1 | 2 |
| Set attribute | 1 | 2 |
| Manage view rectangle | 1 | 2 |
| Visitor | 1 | 6 |

**Table 4.** Summary of the results of the dynamic analysis experiment.

---

[8] Available from `http://star.itc.it/dynamo/` under GPL.

Among the concepts in the lattice, 11 satisfied the crosscutting conditions (scattering and tangling), described in Section 3, for the use-case specific concepts, while 56 (including the 11 above) satisfied the conditions for the generic concepts. Next, both the use-case specific and generic concepts were revisited manually, to determine which ones could be regarded as plausible seeds and which ones should be considered false positives. The criterion followed in this assessment was the following: a concept satisfying the crosscutting conditions is considered a seed if

- it can be associated to a single, well-identified functionality (this usually accounts for the possibility to give it a short description that labels it), and
- some of the classes involved in such a functionality have a different primary responsibility (indicating crosscutting with respect to the principal decomposition).

Of course, due to the nature of crosscutting concerns and the related design decisions, some level of subjectivity still remains (as is the case for the other techniques).

In the end, the list of candidate seeds shown in Table 4 was obtained. The four topmost concerns are use-case specific. As apparent from the second column of the table, and as was the case for the identifier analysis experiment, some crosscutting concerns were detected by multiple concepts. In total, among the 56 generic concepts satisfying the crosscutting conditions, 24 concepts were judged to be associated with 18 crosscutting concerns.

The methods associated with each candidate seed (counted in the last column of Table 4) are indicative of the "aspectizable" functionality. Although they may be not the complete list (dynamic analysis is partial) and may contain false positives, they represent a good starting point for a refactoring intervention aimed at migrating the application to AOSD.

## 5 Comparing the results

In this section we discuss some selected concerns that were identified by the different techniques. We selected some concerns that were detected by all three techniques, as well as a representative set of concerns that were detected by some techniques but not by others. This allows us to clearly pinpoint the strengths and weaknesses of each individual technique.

### 5.1 Selected concerns

Table 5 summarises the concerns we selected. The first column names the concern. The other columns show by what technique(s) the concern was discovered: if a technique discovered the concern, we put a + sign in the corresponding column, otherwise a - sign is in the table.

| Concern | Fan-In Analysis | Identifier Analysis | Dynamic Analysis |
|---|---|---|---|
| Observer | + | + | + |
| Undo | + | + | + |
| Persistence | + | + | + |
| Consistent behavior / Contract enforcement | + | - | - |
| Command execution | + | + | + |
| Bring to front / Send to back | - | - | + |
| Manage handles | - | + | + |
| Move Figures | + (discarded) | + | + |

**Table 5.** A selection of detected concerns in JHotDraw.

**Observer** The Observer design pattern is an example of a concern reported by all techniques. Other examples include *Command execution*, *Undo* functionality and *Persistence*, whose implementation in JHotDraw is described in [6]. Their identification should come as no surprise, because they correspond to well-known aspects, frequently mentioned in AOSD literature, or to functionalities for which an AOSD implementation looks quite natural.

Concerns identified by all three techniques are probably the best starting point for migrating a given application to AOSD, because developers can be quite confident that the concern is very likely to be an aspect. However, the fact that only four of such concerns were discovered, stresses the need for an approach that combines the strengths of different techniques.

**Contract enforcement / consistent behavior** The *contract enforcement* and *consistent behavior* concerns [16] generally describe common functionality required from, or imposed on, the participants in a given context, such as a specific pre-condition check on certain methods in a class hierarchy. An example from the JHotDraw case is the *Command* hierarchy for which the *execute* methods contain code to ensure the pre-condition that an 'active view' reference exists (is not null).

We classify these concerns as a combination of contract enforcement and consistent behavior since these types often have very similar implementations, and choosing a particular type depends mainly on the context and on (personal) interpretation.

Fan-in analysis is particularly suited to address this kind of scattered, cross-cutting functionalities, which involve a large number of calls to the same method, while the other two techniques potentially miss it. In fact, contract enforcement and consistent behavior are usually associated with method calls that occur in *every* execution scenario, so that they cannot be discriminated by any specific use-case. On the other hand, identifier analysis will miss those cases where the methods that enforce a given contract or ensure consistent behavior do not share a common naming scheme.

**Command execution** This concern deals with the executability and the actual execution of objects whose class belongs to the *Command* hierarchy. Identifier analysis identified a concept which contains exactly the *execute* methods in the *Command* hierarchy. Dynamic analysis identified the classes containing *isExecutable* methods. Indeed, the *execute* methods all have the same name and manual inspection showed they exhibit similar behavior: they nearly all make a super call to an *execute* method, invoke a *checkDamage* method and (though not always) invoke a *setUndoActivity* and *getUndoActivity* method. A similar argument can be made for *isExecutable*.

Hence, whereas identifier and dynamic analysis may not detect the more generic Contract enforcement / Consistent behavior aspect directly, they can identify some locations (pointcuts) where potentially such an aspect could be introduced.

**Bring to front / Send to back** The functionality associated with this concern consists of the possibility to bring figures to the front or send them to the back of an image. When exercised, it executes specific methods that have a low fan-in, hence they were not detected by fan-in analysis. Identifier analysis also missed them, because there were not enough methods with a sufficiently similar name to surpass the threshold. Hence, dynamic analysis is the only technique that identified this concern. This example is a good representative of crosscutting concerns that are reported only by dynamic analysis: whenever the methods involved in a functionality are not characterized by a unifying naming scheme (or there are not enough of them), neither do they have high fan-in, the other two techniques are likely to fail.

**Manage handles** A crosscutting functionality is responsible for managing the handles associated with the graphical elements. Such handles support interactive operations, such as resizing of an element, conducted by clicking on the handle and dragging the mouse. This seed is interesting because it is detected by dynamic analysis and by identifier analysis, but in different ways. Identifier analysis detects this concern based on the presence of the word 'handle' in identifiers. Consequently, it misses methods such as `north()`, `south()`, `east()`, `west()`, which are clearly related to this concern, but do not share the lexicon with the others. On the other hand, dynamic analysis reports both the latter methods and (some of) those containing the word 'handle'. However, since not all possible handle interactions have been exercised, the output of dynamic analysis is partial and does not include all the methods reported by identifier analysis.

The *manage handles* concern was missed by the fan-in analysis because the calls are too specific: they are similar but different calls instead of one single called method with a high fan-in.

**Moving figures** The three techniques discard concerns on different bases: some of the concerns are filtered automatically while others are excluded manually.

The *move figures* concern, seeded by the *moveBy* method in the *Figure* classes, is one example where different, subjective decisions can be made depending on whether the concept is classified either as a candidate aspect or as part of the principal decomposition. The *moveBy* methods allow to move a figure with a given offset. The team which used fan-in analysis argued that the original design seems to consider this functionality as part of a *Figure*'s core logic. The other two teams considered it as part of a crosscutting functionality and included it in the list of reported seeds.

This example highlights the difficulty of deciding objectively on what is and what is not an aspect and corroborates our choice to conduct a qualitative, instead of a quantitative, comparison.

## 5.2   Limitations

As a consequence of applying each technique to the same case, some of the limitations of the respective techniques have become obvious. For example, we obtained a better idea of potential 'false negatives', i.e. concerns that were not identified by a particular technique but that were identified by another. Below, we summarise some of the discovered limitations. In the next section we then describe how to partly overcome these limitations by combining different techniques.

**Fan-in analysis** mainly addresses crosscutting concerns that are largely scattered and that have a significant impact on the modularity of the system. The downside of this characteristic is that concerns with a small code footprint and thus with low fan-in values associated, will be missed. For example, the identification of *Observer* design pattern instances is dependent on the number of classes implementing the observer role. These classes contain calls to specific methods in the *subject* class for registering as listeners to the subject's changes. The number of observer classes will determine to a large extent the number of calls to the registration method in the subject role. A collateral effect is the anticipated unsuitability of the technique for analysing small case studies.

**Identifier analysis** tends to produce a lot of detailed results. However, these results typically contain too much noise (false positives), so a more effective filtering of the discovered concepts, as well as of the elements inside those concepts, is needed. In addition, the discovered concepts are often incomplete, in the sense that they do not completely "cover" an aspect or crosscutting concern. Often, more than one concept is needed to describe a single concern, as was the case for the *Observer* aspect. The individual concepts themselves may also need to be completed with additional elements that are not contained in those concepts. This was the case for the *Undo* aspect: in addition to the methods with 'undo' or 'undoable' in their name, some of the methods calling these undo methods need to be considered as part of the core *aspect* as well.

**Dynamic analysis** is partial (i.e., not all methods involved in an aspect are retrieved), being based on specific executions, and it can determine only aspects that can be discriminated by different execution scenarios (e.g., aspects that are exercised in every program execution cannot be detected). Additionally, it does not deal with code that cannot be executed (e.g., code that is part of a larger framework, but that is not used in a specific application).

## 5.3   Complementarity

The three proposed techniques address symptoms of crosscutting functionality, such as scattering and tangling, in quite different ways. As shown in Table 6, fan-in analysis and dynamic analysis show largely complementary result sets: among the 30 concerns identified by either dynamic or fan-in analysis, only 4 are identified by both techniques. This is an expected result. Fan-in analysis focuses on identifying those methods that are called at multiple places. However, when a method is called many times, it is likely to occur in most (if not all) execution traces. Hence, no specific use-case can be defined to isolate the associated functionality, and dynamic analysis will fail to identify it as a seed.

Identifier analysis is the least discriminating of the three techniques and has a large overlap with the other two techniques. When a concern can be identified through fan-in analysis and/or dynamic analysis, identifier analysis can often isolate it too, since a common lexicon is often used in the names of the involved methods.

In the next section, we will use these observations to propose a new aspect mining technique that is a clever combination of the three individual techniques.

| Technique | Concerns |
|---|:---:|
| Dynamic analysis | 18 |
| Fan-in analysis | 16 |
| Dynamic analysis $\bigcup$ Fan-in analysis | 30 |
| Dynamic analysis $\bigcap$ Fan-in analysis | 4 |

**Table 6.** Concerns identified by either dynamic or fan-in analysis.

## 6   Toward interesting combinations

Based on the discussion in the previous section, this section presents three combined aspect mining techniques and reports on the results of applying these combined techniques on the JHotDraw application. Based on the analysis indicators of *recalled methods* and *seed quality* we compare whether these combined techniques provide a more complete coverage of the detected concerns than each of the original techniques individually.

## 6.1   Motivation

As has been explained in the previous sections, the fan-in analysis and dynamic analysis techniques are largely complementary, and address different symptoms of crosscutting. An obvious and interesting combination of these techniques thus consists of simply applying each technique individually and taking the union of the results. Additionally, the seeds in the intersection of the results (if any) are likely to represent the best aspect candidates, because both techniques identify them. This was illustrated in our experiment, in which both techniques identified the *Observer, Undo, Persistence* and *Command execution* candidates.

As for other combinations of the techniques, two interesting observations were considered. First, the manual intervention required by identifier analysis is very time-consuming and is not justified by the fact that it produces more interesting results. This makes the technique less suited than the others for large(r) cases. Second, both fan-in analysis and dynamic analysis identify only candidate seeds that serve as a starting point for seed expansion. Dynamic analysis in particular suffers from this problem as it is based on a (necessarily partial) list of execution scenarios. Similarly, fan-in analysis is only focused on invocations of high fan-in methods, which represent just a portion of the whole concern. Interestingly, while performing fan-in analysis and dynamic analysis, we observed that the classes and methods in the seed expansion often exhibited similar identifiers.

Consequently, we believe better results can be obtained if we use identifier analysis as a seed expansion technique for the seeds identified by either fan-in analysis or dynamic analysis, or by the seeds identified by both these techniques. In this way, the search space for identifier analysis is reduced significantly, and more automation is provided for the manual seed expansion needed by both fan-in analysis and dynamic analysis. A final manual refinement step is anyway necessary, since the expanded seeds may contain false positives and negatives.

In the remainder of this section, we will present three different techniques: a combination of fan-in analysis with identifier analysis, of dynamic analysis with identifier analysis, and of the union of fan-in analysis and dynamic analysis with identifier analysis.

## 6.2   Definition of the combined techniques

The combined techniques work as follows:

1. Identify interesting candidate seeds by applying fan-in analysis, dynamic analysis or both to the application;
   - For candidate seeds identified by dynamic analysis, (manually) filter out those methods that do not pertain to the concern;
2. For each method in the candidate seed, find its enclosing class, and compute the identifiers occurring in the method and the class name, according to the algorithm used by identifier analysis;
3. Apply identifier analysis to the application, and search for a concept, among the concepts it reports, that is "nearest". The nearest concept is the concept that contains most of the identifiers generated in the previous step. If more than one nearest concept exists, take the union of all their elements.

4. Add the methods contained in the nearest concept(s) to the candidate seed.
5. Revise the expanded list of candidate seeds manually to remove false positives and add missing seeds (false negatives).

In what follows, we experimentally validate these techniques on the JHot-Draw case.

## 6.3    Analysis indicators

Before applying the combined techniques, we define two measures to validate the results. The goal is to measure how identified seeds change in terms of precision and recall. Unfortunately, this requires information about all crosscutting concerns present in the application, and this is not available. Therefore, we have chosen alternative metrics, which we call *recalled methods* and *seed quality*.

**Recalled methods** is the number of methods reported in a seed that actually belong to the crosscutting concern.

**Seed quality** is the percentage of a seed's recalled methods with respect to the total number of methods in the seed. This indicator estimates how difficult it is to spot a concern in the methods provided by the seed.

With respect to the definitions above, it is important to remark that for fan-in, two interpretations of seeds are possible: the first takes only the callees with high fan-in into account; the second interpretation includes, besides the callees with high fan-in, also all callers to these methods. These differences stem from the fact that the fan-in technique is actually based on the call-*relation* and the interpretations use either one or both sides of the relation in seed representations. During exploration these differences aren't that important because we can easily navigate from caller to callee and vise versa. However, when we start assessments based on counting elements, these interpretations do have considerable impact.

In the first case, the number of recalled methods will be low (since call-sites are not considered in the seeds), and the seed quality will always be 100% since the high fan-in callees belong to the concern by definition. The second interpretation will result in higher values of recall and yields a more complete picture of the concern. However lower values for seed quality are possible since not all calls may be caused by a crosscutting concern.

Section 6.4 describes the results of applying combined techniques on the JHotDraw appication, and evaluates the above indicators before and after the experiment. We include results for both interpretations of fan-in seeds discussed above.

## 6.4    Experimental results

Table 7 shows the values of the indicators before and after the completion experiment (based on the first interpretation of seeds for fan-in). Although the completion technique can be applied to all concerns identified by either fan-in analysis or dynamic analysis, we performed the experiment only on the concerns identified by all three techniques. The sole reason is that we need to assess how the completion technique influences the recalled methods and seed quality indicators as compared to their initial values, which can only be done for the *Undo, Command execution, Persistence* and *Observer* concerns.

| Concerns | Undo | | Command execution | |
|---|---|---|---|---|
| **Technique** | **Recalled Methods**[*] | **Seed Quality**[*] | **Recalled Methods**[*] | **Seed Quality**[*] |
| Dynamic analysis | 23 | 64% | 20 | 80% |
| Fan-in analysis | 3 | 100% | 3 | 100% |
| Dyn $\bigcup$ Fan-in | 24 | 63% | 22 | 81% |
| Dyn + Identifier | 183 | 55% | 132 | 80% |
| Fan-in + Identifier | 94 | 100% | 132 | 80% |
| (Dyn $\bigcup$ Fan-in) + Identifier | 183 | 55% | 132 | 80% |
| **Concerns** | **Persistence** | | **Observer** | |
| **Technique** | **Recalled Methods**[*] | **Seed Quality**[*] | **Recalled Methods**[*] | **Seed Quality**[*] |
| Dynamic analysis | 29 | 97% | 3 | 100% |
| Fan-in analysis | 6 | 100% | 10 | 100% |
| Dyn $\bigcup$ Fan-in | 32 | 97% | 13 | 100% |
| Dyn + Identifier | 104 | 100% | 121 | 14% |
| Fan-in + Identifier | 104 | 100% | 146 | 15% |
| (Dyn $\bigcup$ Fan-in) + Identifier | 104 | 100% | 146 | 15% |

**Table 7.** Recalled methods and seed quality before and after completion ([*]based on the first interpretation of seeds for fan-in)

When looking at the common results, it is important to note that fan-in seeds point to distinct crosscutting concerns *sorts* that can occur as parts of more complex structures like implementations of the Observer pattern [18, 19]. In the experiments, these are grouped to obtain the same level of granularity obtained by the other techniques.

A deeper look into the results of the completion with identifier analysis reveals interesting information: For the *Undo* concern, the results of both fan-in analysis and dynamic analysis improve a lot in terms of recalled methods (from 23 and 3 up to 183 and 94). There is a negative impact on the seed quality for (completed) dynamic analysis (from 64% down to 55%), but the seed quality for fan-in plus identifier analysis remains at 100%. For the *Command execution* and *Persistence* concerns, the number of recalled methods increases significantly for the completion technique (from 20 and 3 up to 132 and from 29 and 6 up to 104), while the seed quality remains at the same level.

For the *Observer* concern, the results are less encouraging than for the other concerns. Even though the number of recalled methods increases for the completion technique, the quality of the seeds drops to an unacceptable level (from

| Seed | Recalled Methods | Seed Quality |
|------|------------------|--------------|
| Undo (callee #1) | 24 | 92% |
| Undo (callee #2) | 25 | 88% |
| Undo (callee #3) | 24 | 83% |
| Undo (combined) | 73 | 88% |
| Observer (combined) | 83 | 100% |

**Table 8.** Recalled methods and seed quality for fan-in analysis based on the second interpretation of seeds for fan-in

100% down to 14% and 15%). Clearly, the completion does not provide a good expansion of the original seeds. Closer inspection reveals that no clearly distinctive naming convention has been used to implement the *Observer* concern. The *Undo, Command execution* and *Persistence* concerns employ distinctive identifiers such as `undo/undoable, execute/command` and `store/storable`, which are used extensively only within the concern implementation. Consequently, the completion provided by identifier analysis gives good seed expansions. However, the identifiers used for the *Observer* concern are the more general `figure/update/...` that are used extensively in throughout the application, and not only in the concern implementation. Therefore, identifier analysis is not able to provide a good expansion for the seeds found by the other techniques.

An overview of results based on the second interpretation of seeds for fan-in, i.e. taking also the call-sites into account, is shown in Table 8. For the *Undo* concern, we show both the individual values for each of the three high fan-in callees reported as seeds earlier and the recall and seed quality of the combination of these three. The seed quality is lower than 100% in these cases since some of the calls found were not considered to be part of the actual crosscutting concern. For the *Observer* concern we only show the value for the combined high fan-in callees since it would go too far to go over all individual values here. The seed quality is 100% in these cases since there are no calls from outside this concern to the reported callees.

For a detailed discussion of these measurements and an assessment of various quality metrics, we refer to [20] and the fan-in website[9].

## 7 Summary and future work

The purpose of the paper was to compare three different aspect mining techniques, discuss their respective strengths and weaknesses by applying them to a common benchmark application, and develop combined techniques based on this discussion.

---

[9] http://swerl.tudelft.nl/amr/

We observed that all three techniques were able to identify seeds for well-known crosscutting concerns, but that interesting differences arose for other concerns. These differences are largely due to the different ways in which the techniques work. Fan-in analysis is good at identifying seeds that are largely scattered throughout the system and that involve a lot of invocations of the same method, but it cannot be used to analyse smaller applications. Identifier analysis is able to identify seeds when the associated methods have low fan-in, but only if these methods share a common lexicon. The main drawback of this technique is the large number of reported seeds that had to be inspected manually. Finally, dynamic analysis is able to find seeds in the absence of high fan-in values and common identifiers, but the technique is only partial because it relies on execution traces.

We also observed that the three techniques are quite complementary: fan-in analysis and dynamic analysis require a manual effort to expand the seeds into full concerns, whereas identifier analysis covers a large part of a concern, but requires extensive filtering of the reported seeds. Hence, to improve automation of both fan-in analysis and dynamic analysis, and to reduce the search space for identifier analysis, we proposed a combined technique in which seeds from either fan-in analysis or dynamic analysis are expanded automatically by applying identifier analysis. To verify the performance of this combined technique, we applied it to JHotDraw and interpreted the results in terms of two indicators: *recalled methods* and *seed quality*. The measures show that for three out of the four concerns we considered, the combined technique outperforms the individual techniques. In only one case, the combined technique performed worse.

Future work mainly consists of extending our comparison with other aspect mining techniques, and potentially proposing new interesting combinations with such techniques. This will not only allow us to come up with better (combined) aspect mining techniques, but will also allow us to evaluate the three considered techniques even better, as new concerns will be identified that we were not aware of. Additionally, we could come up with extra quality indicators that complement the *recalled methods* and *seed quality* indicators, and empirically establish their validity by considering other benchmark applications as well.

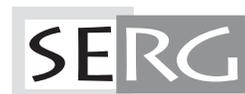